\documentclass[prl,twocolumn,showpacs,superscriptaddress]{revtex4}

\usepackage{amsmath,amssymb}
\usepackage{verbatim}
\usepackage{graphicx}

\usepackage[
      colorlinks=true,
      linkcolor=blue,
      urlcolor=cyan,
      filecolor=magenta,
      citecolor=red,
      pdfstartview=FitV,
      pdftitle={},
        pdfauthor={Arjun Bagchi, Mirah Gary, Zodinmawia},
        pdfsubject={},
        pdfkeywords={},
        pdfpagemode=None,
        bookmarksopen=true
      ]{hyperref}

\DeclareFontFamily{OT1}{rsfs}{}
\DeclareFontShape{OT1}{rsfs}{m}{n}{ <-7> rsfs5 <7-10> rsfs7 <10->rsfs10}{} 
\DeclareMathAlphabet{\mycal}{OT1}{rsfs}{m}{n}

\renewcommand{\L}{{\mathcal{L}}}
\newcommand{\bL}{\bar{{\mathcal{L}}}}

\renewcommand{\a}{\alpha}
\renewcommand{\b}{\beta}

\newcommand{\be}[1]{ \begin{equation}\label{#1} }
\newcommand{\ee}{\end{equation}}
\newcommand{\bea}[1]{\begin{eqnarray}\label{#1} }
\newcommand{\eea}{\end{eqnarray}}
\newcommand{\p}{\partial}
\newcommand{\D}{\Delta}

\newcommand{\refb}[1]{(\ref{#1})}
\newcommand{\<}{\langle}
\renewcommand{\>}{\rangle}

\newcommand{\vev}[1]{\ensuremath{\left\langle#1\right\rangle}}

\def\beaa{\begin{eqnarray*}}
\def\eeaa{\end{eqnarray*}}
\def\bee{\begin{equation*}}
\def\eee{\end{equation*}}
\def\ba{\begin{align}}
\def\ea{\end{align}}

\def\vec{\overrightarrow}
\def\G{\mathcal{G}}

\begin{document}

\title{The BMS Bootstrap}

\author{Arjun Bagchi}
\email{abagchi@iitk.ac.in}
\affiliation{Indian Institute of Technology Kanpur, Kalyanpur, Kanpur 208016. INDIA}

\author{Mirah Gary}
\email{gary@hep.itp.tuwien.ac.at}
\affiliation{Institute for Theoretical Physics, Vienna University of Technology, A-1040 Vienna, AUSTRIA.}

\author{Zodinmawia}
\email{zodin@iitk.ac.in}
\affiliation{Indian Institute of Technology Kanpur, Kalyanpur, Kanpur 208016. INDIA}

\date{\today}

\preprint{}

\begin{abstract} 
We initiate a study of the bootstrap programme for field theories with BMS symmetry. Specifically, we look at two-dimensional field theories with BMS$_3$ symmetry and using highest weight representations we construct the BMS bootstrap equation by formulating the notion of crossing symmetry in the four-point functions of these field theories. In the limit of large central charge, we find analytic expressions for the BMS blocks that are the basic ingredients for the solution of the bootstrap equation. This constitutes, to the best of our knowledge, the first example of the formulation and significant steps towards the solution of a bootstrap equation in a theory which is not a relativistic conformal field theory. 
\end{abstract}


\maketitle

\paragraph{Introduction.}
Bootstrapping is a process that is self-generating or self-sustaining. Historically, it meant an absurd or impossible action, but has come to be used in a much more positive light in the modern world. Much of physics is the study of symmetry and symmetry principles. A particularly useful symmetry, that has found very wide ranging applications starting from the study of phase transitions in statistical mechanics to the use of worldsheet techniques in string theory, is conformal symmetry \cite{DiFrancesco:1997nk}. Field theories with conformal symmetry, or conformal field theories (CFTs) enjoy more symmetry than usual relativistic field theories. In any general dimensions, the relativistic conformal group consists of the Poincare group (rotations, boosts and translations) along with scalings and special conformal transformations. By repeated use of conformal invariance, it is possible to constrain the form of correlation functions, the observables of a CFT, completely. This non-perturbative method of constraining and hence solving CFTs is known as the conformal bootstrap programme \cite{Ferrara:1973yt, Polyakov:1974gs} .  

In $D=2$, the above mentioned finite conformal group is enhanced to two copies of the infinite dimensional Virasoro algebra \cite{Belavin:1984vu} given by:
\be{Vir}
[\L_n, \L_m]= (n-m) \L_{n+m} + \frac{c}{12} \delta_{n+m, 0} (n^3 - n)
\ee
and a second copy $\bL_n$ that commutes with $\L_n$. The power of the infinite algebra in 2D was used to find a class of exact solutions called the minimal models. One of the main ideas behind this was the conformal bootstrap \cite{Belavin:1984vu, Zamolodchikov:1990ww}. It was assumed that there exists an associative operator algebra and this led to a powerful set of constraints on the observables of the theory. This technique helped solve the models of the minimal series, which included the Ising model, as well as the Liouville theory \cite{Zamolodchikov:1995aa}. The power of infinite symmetry, specific to $D=2$, was useful in all of this and the developments of conformal bootstrap techniques remained confined to $D=2$ for quite a while. Recently, following \cite{Dolan:2000ut, Rattazzi:2008pe} and aided by numerical studies, there has been an explosion in activity in generalising the conformal bootstrap programme to dimensions higher than two. The interested reader is referred to \cite{Poland:2016chs, Simmons-Duffin:2016gjk} for a review of the current status in the field. 

\paragraph{BMS symmetry.}
It is obviously of interest to ask whether one can extend the methods of bootstrap to field theories with symmetry structures other than conformal symmetry. In this paper, we initiate a programme of what we call the BMS bootstrap. We will consider field theories invariant under the Bondi-Metzner-Sachs (BMS) group and, drawing inspiration from techniques in CFTs, use symmetry to constrain such theories. In a gravitational theory, the Asymptotic Symmetry Group (ASG) formally captures the symmetries of the theory at infinity. The states of the theory form representations of the ASG which also dictates the symmetry structure of any putative holographically dual field theory living on the boundary of the gravitational theory. The BMS group arises as the ASG of asymptotically flat spacetimes at their null boundary \cite{Bondi:1962, Sachs:1962wk}. For 3D Minkowski spacetimes, the ASG is the BMS$_3$ group, the associated algebra of which is given by \cite{Ashtekar:1996cd, Barnich:2006av} 
\bea{bms3}
&& [L_n, L_m] = (n-m) L_{n+m} + {c_L}\delta_{n+m, 0} (n^3 - n) \crcr
&& [L_n, M_m] = (n-m) M_{n+m} + {c_M}\delta_{n+m, 0} (n^3 - n), \crcr
&& [M_n, M_m] = 0.
\eea
Here $L_n$'s are diffeomorphisms of the circle at null infinity called superrotations, while $M_n$'s are angle dependent translations of the null direction known as supertranslations. \refb{bms3} is also the symmetry algebra of any putative dual 2d field theory which lives on the null boundary of 3d flat space \cite{Bagchi:2010eg}. Constructing the notion of holography for 3d flat spacetimes using the BMS group has been pursued recently with some successes \cite{Bagchi:2012xr, Barnich:2012xq, Barnich:2012aw, Bagchi:2012yk, Barnich:2012rz, Bagchi:2014iea, Bagchi:2015wna}. The reader is referred to \cite{Bagchi:2016bcd} that contains a summary of current research in this direction.  

Our goal in the present paper would be to consider 2d field theories invariant under \refb{bms3} and constrain their properties using self-consistency requirements mimicking the conformal bootstrap programme mentioned previously. Interestingly, field theories with these symmetries arise in many different contexts, e.g. in non-relativistic conformal systems \cite{Bagchi:2009my, Bagchi:2009pe}, on the worldsheet of tensionless string theory \cite{Bagchi:2013bga, Bagchi:2015nca}. Especially in the non-relativistic context, these field theories, also known as Galilean Conformal Field Theories (GCFTs), are expected to play the same role as CFTs and are expected to govern the physics of the fixed points in the renormalisation group flows in Galilean field theories. So the methods and results of this work would have far reaching consequences in many diverse fields. We would also like to emphasise that, to the best of our knowledge, this is the first construction of a bootstrap programme outside the ambit of relativistic conformal field theories.

\paragraph{OPE and recursion relation.}
We will concentrate on 2d field theories with \refb{bms3} as their symmetry algebra. We would also be primarily working in what we call the ``plane" representation. In a field theory with a non-compact spatial direction $x$ and time direction $t$, the vector fields which generate \refb{bms3} in this representation are given by: 
\be{}
L_n = - x^{n+1} \p_x - (n+1) x^n t \p_t, \quad M_n = x^{n+1} \p_t
\ee
We will consider highest weight representations of the algebra \refb{bms3}. This means that the states of the theory (here for the BMS invariant field theory we assume a state-operator correspondence: $\phi(0,0)|0\> = |\phi\>$) are labelled by \cite{Bagchi:2009ca,Bagchi:2009pe}
\be{}
L_0 |\D, \xi\> = \D|\D, \xi\>, \quad M_0 |\D, \xi \> = \xi |\D, \xi\>
\ee
We construct the BMS modules by acting with raising operators $L_{-n}, M_{-n}$ on BMS primary states that are defined as
\be{}
L_n |\D, \xi \> = M_n |\D, \xi \> = 0 \quad \forall n>0.
\ee
For primary operators, the co-ordinate dependence of the 2 and 3-point functions is completely fixed by invariance under the 3d Poincare subgroup (which by abuse of language, we will call the global sub-group) of the BMS$_3$ group generated by $L_{0,\pm1},M_{0,\pm1}$ {\footnote{It is instructive to keep in mind that the 3d Poincare group acts on a 2d field theory and hence this is {\em{not}} a usual relativistic QFT.}}:
\bea{eqn:3-pt}
&& \hspace{-0.5cm}\vev{\phi_1(x_1,t_1)\phi_2(x_2,t_2)} = C_{12} \ x_{12}^{-2\Delta_1} e^{2 \xi_1 \frac{t_{12}}{x_{12}}}\delta_{\Delta_1,\Delta_2}\delta_{\xi_1,\xi_2}\nonumber\\
&& \hspace{-0.5cm} \vev{\phi_1(x_1,t_1)\phi_2(x_2,t_2)\phi_3(x_3,t_3)} \\
 && \hspace{0.3cm} =  C_{123} \ x_{12}^{\D_{123}} x_{23}^{\D_{231}} x_{31}^{\D_{312}} e^{-\xi_{123} \frac{t_{12}}{x_{12}}} e^{-\xi_{312}\frac{t_{31}}{x_{31}}}e^{-\xi_{231} \frac{t_{23}}{x_{23}}}.\nonumber
\eea
Here $C_{12}$ is a normalisation which is taken to be $\delta_{12}$. $\D_{ijk} = - (\D_i + \D_j - \D_k)$ and $\xi_{ijk}$ is defined similarly. $C_{123}$ is an arbitrary constant called the structure constant. This is not fixed by symmetry and depends on the details of the field theory under consideration. 

All information about correlation functions are contained in the operator algebra, which gives the operator product expansion (OPE) of two primary fields as summation over all primaries and their descendants. So, in order to know how correlation functions are constrained by BMS symmetries it is enough to study constraints on the OPE. We make the following ansatz for the OPE of two primary fields with weights $(\D_1,\xi_1)$ and $(\D_2,\xi_2)$
\bea{ope}
&& \phi_{1}(x_1,t_1)\phi_{2}(x_2,t_2) = \sum_{p,\vec{k},\vec{q}} \sum_{\a=0}^{K+Q} x_{12}^{\D_{12p}}\,e^{-\xi_{12p}\frac{t_{12}}{x_{12}}}\\
&& \qquad \qquad \times  C_{12}^{p\{\vec{k},\vec{q}\},\a}x_{12}^{K+Q-\a}t_{12}^{\a}\,\phi_{p}^{\{\vec{k},\vec{q}\}}(x_2,t_2). \nonumber
\eea
Here, for vectors $\vec{k}=(k_1, ..., k_r)$ and $\vec{q}=(q_1,..., q_s)$, we use the following notation for the descendants of the primary field $\phi_p$
\bea{}
\phi_{p}^{\{\vec{k},\vec{q}\}}(x,t)&=&\left(L_{-1}^{k_1}L_{-2}^{k_2}...L_{-r}^{k_r}M_{-1}^{q_1}M_{-2}^{q_2}...M_{-s}^{q_s}\phi_p\right)(x,t)\cr
&\equiv& \left(L_{\vec{k}}M_{\vec{q}}\phi_p\right)(x,t),    
\eea
and $K=\sum_ik_i$ and $Q=\sum_lq_l$. The form of the factor $x^{\D_{12p}}\,e^{-\xi_{12p} \frac{t}{x}}$ is fixed by requiring that the OPE gives the correct 2-pt
function, and the term $\sum_{\a=0}^{K+Q} C_{12}^{p\{\vec{k},\vec{q}\},\a}x^{K+Q-\a}t^{\a}$ is to ensure that both sides of the OPE transforms the same way under the action of $L_{0}$. Using the OPE to find the 3-pt functions and comparing it with \refb{eqn:3-pt}, we find $C_{12}^{p\{0,0\},0}\equiv C_{12}^{p}=C_{p12}$. So, we will rewrite:
\be{}
C_{12}^{p\{\vec{k},\vec{q}\},\a}=C_{12}^{p}\b_{12}^{p\{\vec{k},\vec{q}\},\a},
\ee
where by convention $\b_{12}^{p\{0,0\},0}=1$. The coefficients $\b_{12}^{p\{\vec{k},\vec{q}\},\a}$ are calculated by demanding that both sides of \refb{ope} transforms in the same way under $L_m$ and $M_n$. For simplicity, we take $\D_{1}=\D_{2}=\D, \, \ \xi_{1}=\xi_{2}=\xi$. We apply both sides of \refb{ope} on the vacuum and take $(x_1, t_1; x_2, t_2)=(x,t; 0,0)$  to obtain
\be{ope:states}
\phi(x,t)|\D,\xi\> = \sum_p x^{-2\D+\D_{p}}\,e^{(2\xi-\xi_{p})\frac{t}{x}} \sum_{N\geq \a}C_{12}^{p}x^{N-\a}t^{\a}|N,\a\>,
\ee
where the state
\be{}
|N,\a\>=\sum_{\stackrel{ \{\vec{k},\vec{q}\},}{K+Q=N,\,\a\leq N}
}\b_{12}^{p\{\vec{k},\vec{q}\},\a}L_{\vec{k}}M_{\vec{q}}|\D_p,\xi_p\> \nonumber
\ee
is a descendant state at level $N$ in the BMS module:
\be{}
L_{0}|N,\a\>=(\D_{p}+N)|N,\a\>.
\ee
Operating $L_{n}$ on both sides of sides of \refb{ope:states} and equating coefficients of $x^{-2\D+\D_{p}}\,e^{(2\xi-\xi_{p})\frac{t}{x}}x^{K+n-\a}t^{\a}$, we get
\bea{recurL}
L_{n}|N+n,\a\>&=& \left(N+n \a -\D +n \D +\D _p\right)|N,\a\>\cr
&&+\left(n \xi -n^2 \xi -n \xi _p\right)|N,\a -1\>.
\eea
Similarly, we get two other recursion relations
\bea{recurM0}
M_{0}|N,\a\>&=&\xi_{p} |N,\a\>-(\a +1)|N,\a +1\>, \\
M_{n}|N+n,\a\> &=& \left((n-1) \xi +\xi _p\right)|N,\a\> \\ && \qquad \qquad -(\a+1)|N,\a+1\>. \nonumber
\eea
We can use the above equations to recursively find all $\b_{12}^{p\{\vec{k},\vec{q}\},\a}$. In Table \eqref{table1} below, we display the coefficients calculated for level 1 using these recursion relations. A more detailed discussion of the higher order coefficients will be available in \cite{BGZ2}. So, apart from the structure constants and the spectrum of the primary fields, the form of the OPE is completely fixed by symmetry. Hence given the structure constants, spectrum of primaries in the theory and the central charges, we can completely solve the theory, just like in the case of usual CFTs. These dynamical inputs are the only external inputs needed to completely specify a given BMS-invariant field theory. However, any random sets of these dynamical inputs do not constitute a consistent field theory; they have to satisfy a constraint equation given by the bootstrap equation that arises as a condition for the associativity of the operator algebra. 

\begin{table}[ht]
\def\arraystretch{1.5}
\centering
\begin{tabular}{ |c|c| } 
 \hline
$\b_{12}^{p\{1,0\},0}=\frac{1}{2}$  & $\b_{12}^{p\{0,1\},0}=0$ \\
\hline
$\b_{12}^{p\{1,0\},1}=0$  & $\b_{12}^{p\{0,1\},1}=-\frac{1}{2}$ \\
\hline
\end{tabular}
\caption{Coefficients of OPE at level $N=1$.}
\label{table1}
\end{table}

\paragraph{BMS blocks, crossing symmetry and bootstrap.}
Just like in CFTs, the coordinate dependence of 4-pt functions of primaries in a BMS invariant theory is not completely determined by invariance under the Poincare subgroup $\{L_{0,\pm1},M_{0,\pm1}\}$. They depend on an arbitrary functions $\G_{BMS}(x,t)$ of the BMS analogous of the cross ratios  $x$ and $t$ given by 
\be{}
x=\frac{x_{12}x_{34}}{x_{13}x_{24}},\,\,\,\frac{t}{x}=\frac{t_{12}}{x_{12}}+\frac{t_{34}}{x_{34}}-\frac{t_{13}}{x_{13}}-\frac{t_{24}}{x_{24}}.
\ee 
These cross ratios (and consequently $\G_{BMS}(x,t)$) are invariant under the Poincare subgroup. So, the 4-pt function has the form 
\be{eqn:4pt}
\<\prod_{i=1}^{4}\phi_{i}(x_{i},t_{i})\>= P(\{\D_i,\xi_i,x_{ij},t_{ij}\})\G_{BMS}(x,t).
\ee
where $$P(\{\D_i,\xi_i,x_{ij},t_{ij}\}) = \prod_{i,j} x_{ij}^{\sum_{k=1}^{4}\D_{ijk}/3} e^{-\frac{t_{ij}}{x_{ij}} \sum_{k=1}^{4}\xi_{ijk}/3}.$$
We can now do a coordinate transformation such that 
\be{eqn:gt} 
\{(x_i,t_i) \} \rightarrow \{(\infty,0), (1,0), (x,t), (0,0)\}. 
\ee
where $i = 1 \ldots 4$. This is the BMS analogue of the CFT statement that one can fix four points $\{z_i\} \to \{\infty, 1, z, 0\}$ by conformal symmetry. Correspondingly, we define
\bea{}
G_{34}^{21}(x,t)&\equiv&\lim_{x_{1}\rightarrow\infty, t_{1}\rightarrow0}x^{2\D_{1}} e^{-\frac{2\xi_{1}t_{1}}{x_{1}}}\cr
&& \hspace{-0.5cm}\times \<\phi_{1}(x_{1},t_{1})\phi_{2}(1,0)\phi_{3}(x,t)\phi_{4}(0,0)\>,
\eea
that can be expressed in terms of the in and out states:
\be{}
G_{34}^{21}(x,t)=\<\D_{1},\xi_{1}|\phi_{2}(1,0)\phi_{3}(x,t)|\D_{4},\xi_{4}\>.
\ee
Then 4-pt functions in terms of $G_{34}^{21}(x,t)$ are given by
\be{four_point}
\<\prod_{i=1}^{4}\phi_{i}(x_{i},t_{i})\>=P(\{\D_i,\xi_i,x_{ij},t_{ij}\})f(x,t)^{-1}G_{34}^{21}(x,t),
\ee
where 
\bea{}
f(x,t)&=&(1-x)^{\frac{1}{3}(\D_{231}+\D_{234})}x^{\frac{1}{3}(\D_{341}+\D_{342})}\cr
&&\times e^{\frac{t}{3(1-x)}(\xi_{231}+\xi_{234})}e^{-\frac{t}{3x}(\xi_{341}+\xi_{342})}.\nonumber
\eea
The ordering of operators inside the correlator does not matter except for fermions which would introduce a sign change. So we can move the operators around inside the correlators. So, apart from $G_{34}^{21}(x,t)$ we may also define
\be{}
G_{32}^{41}(x,t)=\<\D_{1},\xi_{1}|\phi_{4}(1,0)\phi_{3}(x,t)|\D_{2},\xi_{2}\>.
\ee 
It can be easily seen from their definition that the functions $G_{ij4}^{kl}(x,t)$  are related by the crossing symmetry
\be{crosssym}
G_{34}^{21}(x,t)=G_{32}^{41}(1-x,-t).
\ee
If we use the OPE between the fields $\phi_3$ and $\phi_4$ in $G_{34}^{21}(x,t)$,  we can express them in terms of three-point functions of primary fields and their descendant. More precisely, using the OPE, the function $G_{34}^{21}(x,t)$ can be decomposed as 
\be{}
G_{34}^{21}(x,t)=\sum_{p}C_{34}^{p}C_{12}^{p}A_{34}^{21}(p|x,t),
\ee
where the four-point conformal block $A_{34}^{21}(p|x,t)$ is the sum of all contribution coming from the primary fields $\phi_p$ and its descendant. It is given by
\bea{}
&& A_{34}^{21}(p|x,t) = (C_{12}^{p})^{-1}x^{-\D_{3}-\D_{4}+\D_{p}}\,e^{(\xi_{3}+\xi_{4}-\xi_{p})\frac{t}{x}}\cr
&& \qquad \qquad \times \sum_{N\geq\a}x^{N-\a}t^{\a}\<\D_{1},\xi_{1}|\phi_{2}(1,0)|N,\a\> \cr
&&=	x^{\D_{34p}}\,e^{-\xi_{34p}\frac{t}{x}} \sum_{\{\vec{k},\vec{q}\}}\left(\sum_{\a=0}^{K+Q}\b_{34}^{p\{\vec{k},\vec{q}\},\a}x^{K+Q-\a}t^{\a}\right)\cr
&&\qquad \quad \times \frac{\<\D_{1},\xi_{1}|\phi_{2}(1,0)L_{\vec{k}}M_{\vec{q}}|\D_{p},\xi_{p}\>}{\<\D_{1},\xi_{1}|\phi_{2}(1,0)|\D_{p},\xi_{p}\>}.
\eea
We have already shown that the coefficients $\b_{12}^{p\{\vec{k},\vec{q}\},\a}$ can be calculated recursively using BMS symmetry.  So, the closed form expression of these blocks are completely determined by symmetry. For the function $G_{32}^{41}(x,t)$ we may use the OPE between the fields $\phi_2$ and $\phi_3$ giving us the expansion in terms of another BMS block $A_{32}^{41}(q|x,t)$
\be{}
G_{32}^{41}(x,t)=\sum_{q}C_{32}^{q}C_{41}^{q}A_{32}^{41}(q|x,t).
\ee
The OPE has to consistent in the sense that \refb{crosssym} has to be satisfied after using the OPE to expand both sides in terms of the blocks. This give us the BMS bootstrap equation
\be{btstrp}
\sum_{p}C_{34}^{p}C_{12}^{p}A_{34}^{21}(p|x,t)=\sum_{q}C_{32}^{q}C_{41}^{q}A_{32}^{41}(q|1-x,-t).
\ee
For any consistent BMS invariant field theory, the structure constants and the spectrum of primary operators have to satisfy \refb{btstrp}. Knowing the closed form expressions of the blocks we can solve \refb{btstrp} to find all the possible consistent BMS invariant theories. However, even though the BMS blocks are completely fixed by symmetry, we will only be able to solve them in a simplifying limit which we now go on to discuss. 

\paragraph{BMS blocks for large central charge.} For even dimensional CFTs with $d\geq 4$, a closed form expression for conformal blocks was obtained for scalar operators in \cite{Dolan:2000ut}. For $2d$ CFT, their method gives the global conformal blocks, which is the large central charge limit of the Virasoro conformal blocks. We employ an analogue of this method to obtain what we will call the BMS global blocks $g_{ij}^{kl}(p|x,t)$. 

If we take the asymptotic limit $c_{L},c_{M}\rightarrow\infty$ in \refb{ope} {\footnote{To readers familiar with the limit of the BMS$_3$ algebra from 2d CFT, the $c_M \to \infty$ limit is perhaps confusing. By this, we implicitly mean that we take $c_M/\xi \to \infty$.}},  the leading terms are given by the descendant fields generated by $L_{-1}$ and $M_{-1}$
\bea{eqn:ope_asymptotic}
&&\phi_{3}(x_3,t_3)\phi_{4}(x_4,t_4) =\sum_{p,\{k,q\}}\sum_{\a=0}^{N=k+q} x_{34}^{\D_{34p}}\,e^{-\xi_{34p}\frac{t_{34}}{x_{34}}}\cr
&& \quad \times C_{34}^{p}\,\b_{34}^{p\{k,q\},\a}x_{34}^{k+q-\a}t_{34}^{\a}(L_{-1})^{k}(M_{-1})^{q}\phi_p(x_4,t_4)\cr
&& \quad  + \mathcal{O}\left(\frac{1}{c_L},\frac{1}{c_M}\right)+...
\eea
So, the function $G_{34}^{21}(x,t)$ has an expansion of the form
\be{}
G_{34}^{21}(x,t)= \sum_{p}C_{12}^{p}C_{34}^{p}\,g_{34}^{21}(p|x,t) + \mathcal{O}\left(\frac{1}{c_L},\frac{1}{c_M}\right)+... \nonumber
\ee
where the global BMS blocks $g_{34}^{21}(p|t,x)$ is the large central charge limit of $A_{34}^{21}(p|x,t)$ given by
\bea{eqn:cb2}
g_{34}^{21}(p|x,t) &=& x^{\D_{34p}}\,e^{-\xi_{34p}\frac{t}{x}} \sum_{\{k,q\}} \sum_{\a=0}^{N=k+q}\b_{34}^{p\{k,q\},\a}x^{N-\a}t^{\a} \cr
&&\hspace{-.8cm} \times \frac{\<\D_{1},\xi_{1}|\phi_{2}(1,0)(L_{-1})^{k}(M_{-1})^{q}|\D_{p},\xi_{p}\>}{\<\D_{1},\xi_{1}|\phi_{2}(1,0)|\D_{p},\xi_{p}\>}
\eea
It is possible to find a differential equation for $g_{34}^{21}(p|x,t)$ by using the requirement that both sides of the OPE transforms the same way under the quadratic Casimirs 
\bea{}
&& \mathcal{C}_1 = M_0^2 - M_1 M_{-1}, \crcr && \mathcal{C}_2 = 4L_0M_0- L_{-1}M_1 - L_1M_{-1} - M_1L_{-1} - M_{-1}L_1 \nonumber
\eea
of the global subgroup of BMS group. For simplicity, we choose $\D_{i=1,2,3,4}=\D,\,\xi_{i=1,2,3,4}=\xi$ and denote the blocks for this special case as $g_{\D_{p},\xi_{p}}(x,t)$. Defining
\be{}
h(p|x,t)=x^{2\D}e^{-\frac{2\xi t}{x}}g_{\D,\xi}(p|x,t),
\ee
we find two differential equations corresponding to the two Casimirs
\bea{diffeq1}
&& \left[\p_t^2+\frac{\xi_p^2}{x^2(x-1)}\right]h(p|x,t)=0,\\
&& \left[ x^2 \p_t - (1-\frac{3}{2} x) x t \p_t^2 + (x-1) x^2 \p_x\p_t \right] h(p|x,t) \nonumber \\
&& \qquad \qquad\qquad\qquad \quad =(\D _p - 1) \xi _p \ h(p|x,t). \label{diffeq2}
\eea
Solving \refb{diffeq1}, \refb{diffeq2} using boundary conditions from \refb{eqn:cb2}, for $|x|<1$, we get:
\bea{}
g_{\D,\xi}(p|x,t)&=& 2^{2 \D _p-2}\, \left(1-x\right)^{-1/2}  \exp{\left(\frac{-\xi_p t}{x\sqrt{1-x}} +2\xi \frac{t}{x} \right)} \nonumber\\
&& \quad \times \,  x^{\D _p-2\D} (1+\sqrt{1-x})^{2-2\D _p}.
\eea
The above equation thus gives an explicit closed-form expression for a global BMS block and is one of the central results of this paper.

\paragraph{Conclusions.}  In this paper, we have initiated the BMS bootstrap programme, which is a systematic procedure to constrain field theories with BMS symmetry. We have focussed entirely on 2d field theories with \refb{bms3} as their symmetry algebra and have been inspired by constructions in 2d CFTs in order to set up the BMS bootstrap equation based on the crossing symmetry of 4-pt functions of BMS primary operators. We have then looked at the large central charge limit to obtain closed-form expressions of what we named the global BMS block. As stressed before, to the best of our knowledge, this is the first example of the construction and significant steps towards the solution of a bootstrap equation in a field theory which has symmetries other than relativistic conformal symmetry. Our programme will help us constrain field theories putatively dual to Minkowski spacetimes. 

Through the so-called BMS/GCA correspondence \cite{Bagchi:2010eg}, our analysis and results in this paper also would be applicable for 2d Galilean conformal field theories and hence would help systematically analyse all Galilean invariant field theories in 2d. In particular, it would be very interesting to investigate whether a set of minimal models exist for 2d GCFTs using bootstrap techniques developed here. 

The algebra \refb{bms3} can be obtained as a contraction of two copies of the Virasoro algebra \refb{Vir}. It should be possible to obtain all the above results as limits of the corresponding analyses in 2d CFTs. This is presently being investigated \cite{BGZ2}. This would provide an independent check of the validity of our analysis in this paper. 

There also exists a particularly interesting limit of the algebra \refb{bms3} where the central term $c_M=0$. It is possible to show through an analysis of null vectors that this leads to a consistent truncation of the theory to a chiral CFT \cite{Bagchi:2009pe}. This has been used in formulating a holographically dual theory called Flatspace Chiral Gravity \cite{Bagchi:2012yk}. It is of interest to check the validity of the chiral truncation in terms of the bootstrap programme and this will be reported in \cite{BGZ2}. 

Among the numerous other future directions, construction of the holographic side of the BMS conformal blocks with the flat-space analogues of geodesic Witten diagrams \cite{Hijano:2015zsa} is a very important and interesting project, which we wish to investigate immediately.  

\smallskip

\paragraph{Acknowledgements.}  Discussions with R. Basu, D. Grumiller, R. Gopakumar, P. Raman and N. V. Suryanarayana are gratefully acknowledged. AB thanks MIT  for hospitality during the initial part of this project and the Fulbright Foundation, the Max Planck society and the Department of Science and Technology, India for partial financial support. MG is supported by the FWF project P27396-N27 and Z by the India-Israel joint research project UGC/PHY/2014236.

\bigskip


\begin{thebibliography}{30}%
\makeatletter
\providecommand \@ifxundefined [1]{%
 \@ifx{#1\undefined}
}%
\providecommand \@ifnum [1]{%
 \ifnum #1\expandafter \@firstoftwo
 \else \expandafter \@secondoftwo
 \fi
}%
\providecommand \@ifx [1]{%
 \ifx #1\expandafter \@firstoftwo
 \else \expandafter \@secondoftwo
 \fi
}%
\providecommand \natexlab [1]{#1}%
\providecommand \enquote  [1]{``#1''}%
\providecommand \bibnamefont  [1]{#1}%
\providecommand \bibfnamefont [1]{#1}%
\providecommand \citenamefont [1]{#1}%
\providecommand \href@noop [0]{\@secondoftwo}%
\providecommand \href [0]{\begingroup \@sanitize@url \@href}%
\providecommand \@href[1]{\@@startlink{#1}\@@href}%
\providecommand \@@href[1]{\endgroup#1\@@endlink}%
\providecommand \@sanitize@url [0]{\catcode `\\12\catcode `\$12\catcode
  `\&12\catcode `\#12\catcode `\^12\catcode `\_12\catcode `\%12\relax}%
\providecommand \@@startlink[1]{}%
\providecommand \@@endlink[0]{}%
\providecommand \url  [0]{\begingroup\@sanitize@url \@url }%
\providecommand \@url [1]{\endgroup\@href {#1}{\urlprefix }}%
\providecommand \urlprefix  [0]{URL }%
\providecommand \Eprint [0]{\href }%
\providecommand \doibase [0]{http://dx.doi.org/}%
\providecommand \selectlanguage [0]{\@gobble}%
\providecommand \bibinfo  [0]{\@secondoftwo}%
\providecommand \bibfield  [0]{\@secondoftwo}%
\providecommand \translation [1]{[#1]}%
\providecommand \BibitemOpen [0]{}%
\providecommand \bibitemStop [0]{}%
\providecommand \bibitemNoStop [0]{.\EOS\space}%
\providecommand \EOS [0]{\spacefactor3000\relax}%
\providecommand \BibitemShut  [1]{\csname bibitem#1\endcsname}%
\let\auto@bib@innerbib\@empty
\bibitem [{\citenamefont {Di~Francesco}\ \emph {et~al.}(1997)\citenamefont
  {Di~Francesco}, \citenamefont {Mathieu},\ and\ \citenamefont
  {Senechal}}]{DiFrancesco:1997nk}%
  \BibitemOpen
  \bibfield  {author} {\bibinfo {author} {\bibfnamefont {P.}~\bibnamefont
  {Di~Francesco}}, \bibinfo {author} {\bibfnamefont {P.}~\bibnamefont
  {Mathieu}}, \ and\ \bibinfo {author} {\bibfnamefont {D.}~\bibnamefont
  {Senechal}},\ }\href {\doibase 10.1007/978-1-4612-2256-9} {\emph {\bibinfo
  {title} {{Conformal Field Theory}}}},\ Graduate Texts in Contemporary
  Physics\ (\bibinfo  {publisher} {Springer-Verlag},\ \bibinfo {address} {New
  York},\ \bibinfo {year} {1997})\BibitemShut {NoStop}%
\bibitem [{\citenamefont {Ferrara}\ \emph {et~al.}(1973)\citenamefont
  {Ferrara}, \citenamefont {Grillo},\ and\ \citenamefont
  {Gatto}}]{Ferrara:1973yt}%
  \BibitemOpen
  \bibfield  {author} {\bibinfo {author} {\bibfnamefont {S.}~\bibnamefont
  {Ferrara}}, \bibinfo {author} {\bibfnamefont {A.~F.}\ \bibnamefont {Grillo}},
  \ and\ \bibinfo {author} {\bibfnamefont {R.}~\bibnamefont {Gatto}},\ }\href
  {\doibase 10.1016/0003-4916(73)90446-6} {\bibfield  {journal} {\bibinfo
  {journal} {Annals Phys.}\ }\textbf {\bibinfo {volume} {76}},\ \bibinfo
  {pages} {161} (\bibinfo {year} {1973})}\BibitemShut {NoStop}%
\bibitem [{\citenamefont {Polyakov}(1974)}]{Polyakov:1974gs}%
  \BibitemOpen
  \bibfield  {author} {\bibinfo {author} {\bibfnamefont {A.~M.}\ \bibnamefont
  {Polyakov}},\ }\href@noop {} {\bibfield  {journal} {\bibinfo  {journal} {Zh.
  Eksp. Teor. Fiz.}\ }\textbf {\bibinfo {volume} {66}},\ \bibinfo {pages} {23}
  (\bibinfo {year} {1974})}\BibitemShut {NoStop}%
\bibitem [{\citenamefont {Belavin}\ \emph {et~al.}(1984)\citenamefont
  {Belavin}, \citenamefont {Polyakov},\ and\ \citenamefont
  {Zamolodchikov}}]{Belavin:1984vu}%
  \BibitemOpen
  \bibfield  {author} {\bibinfo {author} {\bibfnamefont {A.~A.}\ \bibnamefont
  {Belavin}}, \bibinfo {author} {\bibfnamefont {A.~M.}\ \bibnamefont
  {Polyakov}}, \ and\ \bibinfo {author} {\bibfnamefont {A.~B.}\ \bibnamefont
  {Zamolodchikov}},\ }\href {\doibase 10.1016/0550-3213(84)90052-X} {\bibfield
  {journal} {\bibinfo  {journal} {Nucl. Phys.}\ }\textbf {\bibinfo {volume}
  {B241}},\ \bibinfo {pages} {333} (\bibinfo {year} {1984})}\BibitemShut
  {NoStop}%
\bibitem [{\citenamefont {Zamolodchikov}\ and\ \citenamefont
  {Zamolodchikov}(1990)}]{Zamolodchikov:1990ww}%
  \BibitemOpen
  \bibfield  {author} {\bibinfo {author} {\bibfnamefont {A.~B.}\ \bibnamefont
  {Zamolodchikov}}\ and\ \bibinfo {author} {\bibfnamefont {A.~B.}\ \bibnamefont
  {Zamolodchikov}},\ }\href@noop {} {\  (\bibinfo {year} {1990})}\BibitemShut
  {NoStop}%
\bibitem [{\citenamefont {Zamolodchikov}\ and\ \citenamefont
  {Zamolodchikov}(1996)}]{Zamolodchikov:1995aa}%
  \BibitemOpen
  \bibfield  {author} {\bibinfo {author} {\bibfnamefont {A.~B.}\ \bibnamefont
  {Zamolodchikov}}\ and\ \bibinfo {author} {\bibfnamefont {A.~B.}\ \bibnamefont
  {Zamolodchikov}},\ }\href {\doibase 10.1016/0550-3213(96)00351-3} {\bibfield
  {journal} {\bibinfo  {journal} {Nucl. Phys.}\ }\textbf {\bibinfo {volume}
  {B477}},\ \bibinfo {pages} {577} (\bibinfo {year} {1996})},\ \Eprint
  {http://arxiv.org/abs/hep-th/9506136} {arXiv:hep-th/9506136 [hep-th]}
  \BibitemShut {NoStop}%
\bibitem [{\citenamefont {Dolan}\ and\ \citenamefont
  {Osborn}(2001)}]{Dolan:2000ut}%
  \BibitemOpen
  \bibfield  {author} {\bibinfo {author} {\bibfnamefont {F.~A.}\ \bibnamefont
  {Dolan}}\ and\ \bibinfo {author} {\bibfnamefont {H.}~\bibnamefont {Osborn}},\
  }\href {\doibase 10.1016/S0550-3213(01)00013-X} {\bibfield  {journal}
  {\bibinfo  {journal} {Nucl. Phys.}\ }\textbf {\bibinfo {volume} {B599}},\
  \bibinfo {pages} {459} (\bibinfo {year} {2001})},\ \Eprint
  {http://arxiv.org/abs/hep-th/0011040} {arXiv:hep-th/0011040 [hep-th]}
  \BibitemShut {NoStop}%
\bibitem [{\citenamefont {Rattazzi}\ \emph {et~al.}(2008)\citenamefont
  {Rattazzi}, \citenamefont {Rychkov}, \citenamefont {Tonni},\ and\
  \citenamefont {Vichi}}]{Rattazzi:2008pe}%
  \BibitemOpen
  \bibfield  {author} {\bibinfo {author} {\bibfnamefont {R.}~\bibnamefont
  {Rattazzi}}, \bibinfo {author} {\bibfnamefont {V.~S.}\ \bibnamefont
  {Rychkov}}, \bibinfo {author} {\bibfnamefont {E.}~\bibnamefont {Tonni}}, \
  and\ \bibinfo {author} {\bibfnamefont {A.}~\bibnamefont {Vichi}},\ }\href
  {\doibase 10.1088/1126-6708/2008/12/031} {\bibfield  {journal} {\bibinfo
  {journal} {JHEP}\ }\textbf {\bibinfo {volume} {12}},\ \bibinfo {pages} {031}
  (\bibinfo {year} {2008})},\ \Eprint {http://arxiv.org/abs/0807.0004}
  {arXiv:0807.0004 [hep-th]} \BibitemShut {NoStop}%
\bibitem [{\citenamefont {Poland}\ and\ \citenamefont
  {Simmons-Duffin}(2016)}]{Poland:2016chs}%
  \BibitemOpen
  \bibfield  {author} {\bibinfo {author} {\bibfnamefont {D.}~\bibnamefont
  {Poland}}\ and\ \bibinfo {author} {\bibfnamefont {D.}~\bibnamefont
  {Simmons-Duffin}},\ }\href {\doibase 10.1038/nphys3761} {\bibfield  {journal}
  {\bibinfo  {journal} {Nature Phys.}\ }\textbf {\bibinfo {volume} {12}},\
  \bibinfo {pages} {535} (\bibinfo {year} {2016})}\BibitemShut {NoStop}%
\bibitem [{\citenamefont {Simmons-Duffin}(2016)}]{Simmons-Duffin:2016gjk}%
  \BibitemOpen
  \bibfield  {author} {\bibinfo {author} {\bibfnamefont {D.}~\bibnamefont
  {Simmons-Duffin}},\ }\href@noop {} {\  (\bibinfo {year} {2016})},\ \Eprint
  {http://arxiv.org/abs/1602.07982} {arXiv:1602.07982 [hep-th]} \BibitemShut
  {NoStop}%
\bibitem [{\citenamefont {Bondi}\ \emph {et~al.}(1962)\citenamefont {Bondi},
  \citenamefont {van~der Burg},\ and\ \citenamefont {Metzner}}]{Bondi:1962}%
  \BibitemOpen
  \bibfield  {author} {\bibinfo {author} {\bibfnamefont {H.}~\bibnamefont
  {Bondi}}, \bibinfo {author} {\bibfnamefont {M.}~\bibnamefont {van~der Burg}},
  \ and\ \bibinfo {author} {\bibfnamefont {A.}~\bibnamefont {Metzner}},\
  }\href@noop {} {\bibfield  {journal} {\bibinfo  {journal} {Proc. Roy. Soc.
  London}\ }\textbf {\bibinfo {volume} {A269}},\ \bibinfo {pages} {21}
  (\bibinfo {year} {1962})}\BibitemShut {NoStop}%
\bibitem [{\citenamefont {Sachs}(1962)}]{Sachs:1962wk}%
  \BibitemOpen
  \bibfield  {author} {\bibinfo {author} {\bibfnamefont {R.~K.}\ \bibnamefont
  {Sachs}},\ }\href {\doibase 10.1098/rspa.1962.0206} {\bibfield  {journal}
  {\bibinfo  {journal} {Proc. Roy. Soc. Lond.}\ }\textbf {\bibinfo {volume}
  {A270}},\ \bibinfo {pages} {103} (\bibinfo {year} {1962})}\BibitemShut
  {NoStop}%
\bibitem [{\citenamefont {Ashtekar}\ \emph {et~al.}(1997)\citenamefont
  {Ashtekar}, \citenamefont {Bicak},\ and\ \citenamefont
  {Schmidt}}]{Ashtekar:1996cd}%
  \BibitemOpen
  \bibfield  {author} {\bibinfo {author} {\bibfnamefont {A.}~\bibnamefont
  {Ashtekar}}, \bibinfo {author} {\bibfnamefont {J.}~\bibnamefont {Bicak}}, \
  and\ \bibinfo {author} {\bibfnamefont {B.~G.}\ \bibnamefont {Schmidt}},\
  }\href {\doibase 10.1103/PhysRevD.55.669} {\bibfield  {journal} {\bibinfo
  {journal} {Phys.Rev.}\ }\textbf {\bibinfo {volume} {D55}},\ \bibinfo {pages}
  {669} (\bibinfo {year} {1997})},\ \Eprint
  {http://arxiv.org/abs/gr-qc/9608042} {arXiv:gr-qc/9608042 [gr-qc]}
  \BibitemShut {NoStop}%
\bibitem [{\citenamefont {Barnich}\ and\ \citenamefont
  {Compere}(2007)}]{Barnich:2006av}%
  \BibitemOpen
  \bibfield  {author} {\bibinfo {author} {\bibfnamefont {G.}~\bibnamefont
  {Barnich}}\ and\ \bibinfo {author} {\bibfnamefont {G.}~\bibnamefont
  {Compere}},\ }\href {\doibase 10.1088/0264-9381/24/5/F01,
  10.1088/0264-9381/24/11/C01} {\bibfield  {journal} {\bibinfo  {journal}
  {Class.Quant.Grav.}\ }\textbf {\bibinfo {volume} {24}},\ \bibinfo {pages}
  {F15} (\bibinfo {year} {2007})},\ \Eprint
  {http://arxiv.org/abs/gr-qc/0610130} {arXiv:gr-qc/0610130 [gr-qc]}
  \BibitemShut {NoStop}%
\bibitem [{\citenamefont {Bagchi}(2010)}]{Bagchi:2010eg}%
  \BibitemOpen
  \bibfield  {author} {\bibinfo {author} {\bibfnamefont {A.}~\bibnamefont
  {Bagchi}},\ }\href {\doibase 10.1103/PhysRevLett.105.171601} {\bibfield
  {journal} {\bibinfo  {journal} {Phys. Rev. Lett.}\ }\textbf {\bibinfo
  {volume} {105}},\ \bibinfo {pages} {171601} (\bibinfo {year} {2010})},\
  \Eprint {http://arxiv.org/abs/1006.3354} {arXiv:1006.3354 [hep-th]}
  \BibitemShut {NoStop}%
\bibitem [{\citenamefont {Bagchi}\ \emph {et~al.}(2013)\citenamefont {Bagchi},
  \citenamefont {Detournay}, \citenamefont {Fareghbal},\ and\ \citenamefont
  {Simon}}]{Bagchi:2012xr}%
  \BibitemOpen
  \bibfield  {author} {\bibinfo {author} {\bibfnamefont {A.}~\bibnamefont
  {Bagchi}}, \bibinfo {author} {\bibfnamefont {S.}~\bibnamefont {Detournay}},
  \bibinfo {author} {\bibfnamefont {R.}~\bibnamefont {Fareghbal}}, \ and\
  \bibinfo {author} {\bibfnamefont {J.}~\bibnamefont {Simon}},\ }\href
  {\doibase 10.1103/PhysRevLett.110.141302} {\bibfield  {journal} {\bibinfo
  {journal} {Phys. Rev. Lett.}\ }\textbf {\bibinfo {volume} {110}},\ \bibinfo
  {pages} {141302} (\bibinfo {year} {2013})},\ \Eprint
  {http://arxiv.org/abs/1208.4372} {arXiv:1208.4372 [hep-th]} \BibitemShut
  {NoStop}%
\bibitem [{\citenamefont {Barnich}(2012)}]{Barnich:2012xq}%
  \BibitemOpen
  \bibfield  {author} {\bibinfo {author} {\bibfnamefont {G.}~\bibnamefont
  {Barnich}},\ }\href {\doibase 10.1007/JHEP10(2012)095} {\bibfield  {journal}
  {\bibinfo  {journal} {JHEP}\ }\textbf {\bibinfo {volume} {1210}},\ \bibinfo
  {pages} {095} (\bibinfo {year} {2012})},\ \Eprint
  {http://arxiv.org/abs/1208.4371} {arXiv:1208.4371 [hep-th]} \BibitemShut
  {NoStop}%
\bibitem [{\citenamefont {Barnich}\ \emph {et~al.}(2012)\citenamefont
  {Barnich}, \citenamefont {Gomberoff},\ and\ \citenamefont
  {Gonzalez}}]{Barnich:2012aw}%
  \BibitemOpen
  \bibfield  {author} {\bibinfo {author} {\bibfnamefont {G.}~\bibnamefont
  {Barnich}}, \bibinfo {author} {\bibfnamefont {A.}~\bibnamefont {Gomberoff}},
  \ and\ \bibinfo {author} {\bibfnamefont {H.~A.}\ \bibnamefont {Gonzalez}},\
  }\href {\doibase 10.1103/PhysRevD.86.024020} {\bibfield  {journal} {\bibinfo
  {journal} {Phys.Rev.}\ }\textbf {\bibinfo {volume} {D86}},\ \bibinfo {pages}
  {024020} (\bibinfo {year} {2012})},\ \Eprint {http://arxiv.org/abs/1204.3288}
  {arXiv:1204.3288 [gr-qc]} \BibitemShut {NoStop}%
\bibitem [{\citenamefont {Bagchi}\ \emph {et~al.}(2012)\citenamefont {Bagchi},
  \citenamefont {Detournay},\ and\ \citenamefont {Grumiller}}]{Bagchi:2012yk}%
  \BibitemOpen
  \bibfield  {author} {\bibinfo {author} {\bibfnamefont {A.}~\bibnamefont
  {Bagchi}}, \bibinfo {author} {\bibfnamefont {S.}~\bibnamefont {Detournay}}, \
  and\ \bibinfo {author} {\bibfnamefont {D.}~\bibnamefont {Grumiller}},\ }\href
  {\doibase 10.1103/PhysRevLett.109.151301} {\bibfield  {journal} {\bibinfo
  {journal} {Phys.Rev.Lett.}\ }\textbf {\bibinfo {volume} {109}},\ \bibinfo
  {pages} {151301} (\bibinfo {year} {2012})},\ \Eprint
  {http://arxiv.org/abs/1208.1658} {arXiv:1208.1658 [hep-th]} \BibitemShut
  {NoStop}%
\bibitem [{\citenamefont {Barnich}\ \emph {et~al.}(2013)\citenamefont
  {Barnich}, \citenamefont {Gomberoff},\ and\ \citenamefont
  {González}}]{Barnich:2012rz}%
  \BibitemOpen
  \bibfield  {author} {\bibinfo {author} {\bibfnamefont {G.}~\bibnamefont
  {Barnich}}, \bibinfo {author} {\bibfnamefont {A.}~\bibnamefont {Gomberoff}},
  \ and\ \bibinfo {author} {\bibfnamefont {H.~A.}\ \bibnamefont {González}},\
  }\href {\doibase 10.1103/PhysRevD.87.124032} {\bibfield  {journal} {\bibinfo
  {journal} {Phys. Rev.}\ }\textbf {\bibinfo {volume} {D87}},\ \bibinfo {pages}
  {124032} (\bibinfo {year} {2013})},\ \Eprint {http://arxiv.org/abs/1210.0731}
  {arXiv:1210.0731 [hep-th]} \BibitemShut {NoStop}%
\bibitem [{\citenamefont {Bagchi}\ \emph {et~al.}(2014)\citenamefont {Bagchi},
  \citenamefont {Basu}, \citenamefont {Grumiller},\ and\ \citenamefont
  {Riegler}}]{Bagchi:2014iea}%
  \BibitemOpen
  \bibfield  {author} {\bibinfo {author} {\bibfnamefont {A.}~\bibnamefont
  {Bagchi}}, \bibinfo {author} {\bibfnamefont {R.}~\bibnamefont {Basu}},
  \bibinfo {author} {\bibfnamefont {D.}~\bibnamefont {Grumiller}}, \ and\
  \bibinfo {author} {\bibfnamefont {M.}~\bibnamefont {Riegler}},\ }\href@noop
  {} {\  (\bibinfo {year} {2014})},\ \Eprint {http://arxiv.org/abs/1410.4089}
  {arXiv:1410.4089 [hep-th]} \BibitemShut {NoStop}%
\bibitem [{\citenamefont {Bagchi}\ \emph
  {et~al.}(2016{\natexlab{a}})\citenamefont {Bagchi}, \citenamefont
  {Grumiller},\ and\ \citenamefont {Merbis}}]{Bagchi:2015wna}%
  \BibitemOpen
  \bibfield  {author} {\bibinfo {author} {\bibfnamefont {A.}~\bibnamefont
  {Bagchi}}, \bibinfo {author} {\bibfnamefont {D.}~\bibnamefont {Grumiller}}, \
  and\ \bibinfo {author} {\bibfnamefont {W.}~\bibnamefont {Merbis}},\ }\href
  {\doibase 10.1103/PhysRevD.93.061502} {\bibfield  {journal} {\bibinfo
  {journal} {Phys. Rev.}\ }\textbf {\bibinfo {volume} {D93}},\ \bibinfo {pages}
  {061502} (\bibinfo {year} {2016}{\natexlab{a}})},\ \Eprint
  {http://arxiv.org/abs/1507.05620} {arXiv:1507.05620 [hep-th]} \BibitemShut
  {NoStop}%
\bibitem [{\citenamefont {Bagchi}\ \emph
  {et~al.}(2016{\natexlab{b}})\citenamefont {Bagchi}, \citenamefont {Basu},
  \citenamefont {Kakkar},\ and\ \citenamefont {Mehra}}]{Bagchi:2016bcd}%
  \BibitemOpen
  \bibfield  {author} {\bibinfo {author} {\bibfnamefont {A.}~\bibnamefont
  {Bagchi}}, \bibinfo {author} {\bibfnamefont {R.}~\bibnamefont {Basu}},
  \bibinfo {author} {\bibfnamefont {A.}~\bibnamefont {Kakkar}}, \ and\ \bibinfo
  {author} {\bibfnamefont {A.}~\bibnamefont {Mehra}},\ }\href@noop {} {\
  (\bibinfo {year} {2016}{\natexlab{b}})},\ \Eprint
  {http://arxiv.org/abs/1609.06203} {arXiv:1609.06203 [hep-th]} \BibitemShut
  {NoStop}%
\bibitem [{\citenamefont {Bagchi}\ and\ \citenamefont
  {Gopakumar}(2009)}]{Bagchi:2009my}%
  \BibitemOpen
  \bibfield  {author} {\bibinfo {author} {\bibfnamefont {A.}~\bibnamefont
  {Bagchi}}\ and\ \bibinfo {author} {\bibfnamefont {R.}~\bibnamefont
  {Gopakumar}},\ }\href {\doibase 10.1088/1126-6708/2009/07/037} {\bibfield
  {journal} {\bibinfo  {journal} {JHEP}\ }\textbf {\bibinfo {volume} {0907}},\
  \bibinfo {pages} {037} (\bibinfo {year} {2009})},\ \Eprint
  {http://arxiv.org/abs/0902.1385} {arXiv:0902.1385 [hep-th]} \BibitemShut
  {NoStop}%
\bibitem [{\citenamefont {Bagchi}\ \emph {et~al.}(2010)\citenamefont {Bagchi},
  \citenamefont {Gopakumar}, \citenamefont {Mandal},\ and\ \citenamefont
  {Miwa}}]{Bagchi:2009pe}%
  \BibitemOpen
  \bibfield  {author} {\bibinfo {author} {\bibfnamefont {A.}~\bibnamefont
  {Bagchi}}, \bibinfo {author} {\bibfnamefont {R.}~\bibnamefont {Gopakumar}},
  \bibinfo {author} {\bibfnamefont {I.}~\bibnamefont {Mandal}}, \ and\ \bibinfo
  {author} {\bibfnamefont {A.}~\bibnamefont {Miwa}},\ }\href {\doibase
  10.1007/JHEP08(2010)004} {\bibfield  {journal} {\bibinfo  {journal} {JHEP}\
  }\textbf {\bibinfo {volume} {1008}},\ \bibinfo {pages} {004} (\bibinfo {year}
  {2010})},\ \Eprint {http://arxiv.org/abs/0912.1090} {arXiv:0912.1090
  [hep-th]} \BibitemShut {NoStop}%
\bibitem [{\citenamefont {Bagchi}(2013)}]{Bagchi:2013bga}%
  \BibitemOpen
  \bibfield  {author} {\bibinfo {author} {\bibfnamefont {A.}~\bibnamefont
  {Bagchi}},\ }\href {\doibase 10.1007/JHEP05(2013)141} {\bibfield  {journal}
  {\bibinfo  {journal} {JHEP}\ }\textbf {\bibinfo {volume} {05}},\ \bibinfo
  {pages} {141} (\bibinfo {year} {2013})},\ \Eprint
  {http://arxiv.org/abs/1303.0291} {arXiv:1303.0291 [hep-th]} \BibitemShut
  {NoStop}%
\bibitem [{\citenamefont {Bagchi}\ \emph
  {et~al.}(2016{\natexlab{c}})\citenamefont {Bagchi}, \citenamefont
  {Chakrabortty},\ and\ \citenamefont {Parekh}}]{Bagchi:2015nca}%
  \BibitemOpen
  \bibfield  {author} {\bibinfo {author} {\bibfnamefont {A.}~\bibnamefont
  {Bagchi}}, \bibinfo {author} {\bibfnamefont {S.}~\bibnamefont
  {Chakrabortty}}, \ and\ \bibinfo {author} {\bibfnamefont {P.}~\bibnamefont
  {Parekh}},\ }\href {\doibase 10.1007/JHEP01(2016)158} {\bibfield  {journal}
  {\bibinfo  {journal} {JHEP}\ }\textbf {\bibinfo {volume} {01}},\ \bibinfo
  {pages} {158} (\bibinfo {year} {2016}{\natexlab{c}})},\ \Eprint
  {http://arxiv.org/abs/1507.04361} {arXiv:1507.04361 [hep-th]} \BibitemShut
  {NoStop}%
\bibitem [{\citenamefont {Bagchi}\ and\ \citenamefont
  {Mandal}(2009)}]{Bagchi:2009ca}%
  \BibitemOpen
  \bibfield  {author} {\bibinfo {author} {\bibfnamefont {A.}~\bibnamefont
  {Bagchi}}\ and\ \bibinfo {author} {\bibfnamefont {I.}~\bibnamefont
  {Mandal}},\ }\href {\doibase 10.1016/j.physletb.2009.04.030} {\bibfield
  {journal} {\bibinfo  {journal} {Phys.Lett.}\ }\textbf {\bibinfo {volume}
  {B675}},\ \bibinfo {pages} {393} (\bibinfo {year} {2009})},\ \Eprint
  {http://arxiv.org/abs/0903.4524} {arXiv:0903.4524 [hep-th]} \BibitemShut
  {NoStop}%
\bibitem [{\citenamefont {Bagchi}\ \emph {et~al.}()\citenamefont {Bagchi},
  \citenamefont {Gary},\ and\ \citenamefont {Zodinmawia}}]{BGZ2}%
  \BibitemOpen
  \bibfield  {author} {\bibinfo {author} {\bibfnamefont {A.}~\bibnamefont
  {Bagchi}}, \bibinfo {author} {\bibfnamefont {M.}~\bibnamefont {Gary}}, \ and\
  \bibinfo {author} {\bibnamefont {Zodinmawia}},\ }\href@noop {} {\ }\Eprint
  {http://arxiv.org/abs/To appear} {To appear} \BibitemShut {NoStop}%
\bibitem [{\citenamefont {Hijano}\ \emph {et~al.}(2016)\citenamefont {Hijano},
  \citenamefont {Kraus}, \citenamefont {Perlmutter},\ and\ \citenamefont
  {Snively}}]{Hijano:2015zsa}%
  \BibitemOpen
  \bibfield  {author} {\bibinfo {author} {\bibfnamefont {E.}~\bibnamefont
  {Hijano}}, \bibinfo {author} {\bibfnamefont {P.}~\bibnamefont {Kraus}},
  \bibinfo {author} {\bibfnamefont {E.}~\bibnamefont {Perlmutter}}, \ and\
  \bibinfo {author} {\bibfnamefont {R.}~\bibnamefont {Snively}},\ }\href
  {\doibase 10.1007/JHEP01(2016)146} {\bibfield  {journal} {\bibinfo  {journal}
  {JHEP}\ }\textbf {\bibinfo {volume} {01}},\ \bibinfo {pages} {146} (\bibinfo
  {year} {2016})},\ \Eprint {http://arxiv.org/abs/1508.00501} {arXiv:1508.00501
  [hep-th]} \BibitemShut {NoStop}%
\end{thebibliography}
\end{document}